# Addressing the Impact of Data Truncation and Parameter Uncertainty on Operational Risk Estimates

**Xiaolin Luo**
CSIRO Mathematical and Information Sciences, Sydney, Locked bag 17, North Ryde, NSW, 1670, Australia. e-mail: Xiaolin.Luo@csiro.au

**Pavel V. Shevchenko**
CSIRO Mathematical and Information Sciences, Sydney, Locked bag 17, North Ryde, NSW, 1670, Australia. e-mail: Pavel.Shevchenko@csiro.au

**John B. Donnelly**
CSIRO Mathematical and Information Sciences, Sydney, Locked bag 17, North Ryde, NSW, 1670, Australia. e-mail: John.B.Donnelly@csiro.au

**Abstract**

Typically, operational risk losses are reported above some threshold. This paper studies the impact of ignoring data truncation on the 0.999 quantile of the annual loss distribution for operational risk for a broad range of distribution parameters and truncation levels. Loss frequency and severity are modelled by the Poisson and Lognormal distributions respectively. Two cases of ignoring data truncation are studied: the "naive model" - fitting a Lognormal distribution with support on a positive semi-infinite interval, and "shifted model" - fitting a Lognormal distribution shifted to the truncation level. For all practical cases, the "naive model" leads to underestimation (that can be severe) of the 0.999 quantile. The "shifted model" overestimates the 0.999 quantile except some cases of small underestimation for large truncation levels. Conservative estimation of capital charge is usually acceptable and the use of the "shifted model" can be justified while the "naive model" should not be allowed. However, if parameter uncertainty is taken into account (in practice it is often ignored), the "shifted model" can lead to considerable underestimation of capital charge. This is demonstrated with a practical example.

# 1. Introduction

The Basel II Accord requires banks to meet a capital requirement for operational risk as part of an overall risk-based capital framework, see BIS (2006). To arrive at a reliable estimate for capital charge, more and more banks have adopted the Advanced Measurement Approaches (AMA) with increasing sophistication and complexity. Under the Loss Distribution Approach (LDA) for AMA, severity and frequency distributions of the loss events are estimated for each risk cell in a bank over a one year time horizon. The industry usually refers these risk cells as "risk nodes" in the Basel II regulatory matrix of eight business lines by seven risk types. The capital charge for operational risk is then based on the 0.999 quantile of the distribution for total annual loss (across all risk cells). If the assumptions and quantification of dependencies between risks are not sound then the 0.999 annual loss quantiles should be added across the risk cells to obtain a conservative estimate of capital charge.

Accurate modelling of the loss severity and frequency distributions is the key to determining a reliable capital charge estimate. There are various aspects to operational risk modelling, see for example Chavez-Demoulin, Embrechts and Nešlehová (2006) or Cruz (2004). One of the challenges in modelling operational risk is the lack of complete data – often a bank's internal data are not reported below a certain level. These data are said to be left-truncated. Generally speaking, missing data increase uncertainty in modelling. Sometimes a threshold level is introduced to avoid difficulties with collection of too many small losses. Industry data, available through external databases from vendors (e.g. OpVar® Database) and consortia of banks (e.g. ORX), are provided above a threshold only. The 2002 LDCE (Loss Data Collection Exercises, conducted by Basel Committee on Banking Supervision) observed that the majority of banks used minimum cut-off levels at or below €10,000, while the rest either used higher cut-off levels or provided no information on the cut-off levels.

Statistically consistent methodology, such as fitting a proper truncated severity distribution using Maximum Likelihood Estimation (see e.g. Frachot, Moudoulaud and Roncalli (2004)) or the Expectation Maximization algorithm (see e.g. Bee (2005)), is required to model missing data correctly. This approach leads to asymptotically unbiased estimates for distribution parameters and is hereafter referred to as the "unbiased model". Unfortunately, the errors in the fitted parameters can be large and the fitting procedure may not be stable or may result in unrealistic parameter estimates, especially for high truncation levels. These problems motivate a modeler to seek simpler alternatives. One of the alternatives is a simplified approach, hereafter referred to as the "naive model", which ignores the missing data and fits standard un-truncated severity distributions (with support on a positive semi-infinite interval) to the truncated data.

This "naive model" will lead to underestimation of capital charge, because it underestimates both loss frequency and severity simultaneously. The quantitative impact of this approach has been analysed in some detail in the literature; see Baud, Frachot and Roncalli (2003), Chernobai *et al.* (2005). The analysis by Chernobai *et al.* (2005) demonstrated that the "naive model" may cause significant underestimation of capital charge in the case of the Lognormal distribution. Another simplified approach, hereafter referred as the "shifted model", is to fit a standard distribution shifted to the truncation level. When a shifted distribution is fitted to data above the truncation level, the recorded data are treated as complete and the missing fraction below truncation is ignored. In this instance all losses are assumed to be above the truncation level and the impact on the capital charge estimate is not obvious – it underestimates loss frequency but overestimates loss severity. In a recent analysis of operational risk losses (collected over many institutions) by Dutta and Perry (2006), the "shifted model" was used to fit data truncated below US$10,000 for all distributions (considered in the study) except the Lognormal. Lognormal distribution was fitted using the "unbiased model" but it was noted that sometimes it lead to unreasonable parameter estimates unless the truncation level was adjusted upwards.

The compound Poisson-Lognormal distribution, which is suggested by the Bank for International Settlements, BIS (2006), and studied by Chernobai *et al.* (2005), is used in the present work to demonstrate the impacts of the above-mentioned simplified approaches. The parameters of the Lognormal distribution under the "shifted" and "naive" models are expressed via the parameters of the true Lognormal distribution in a way consistent with maximum likelihood fitting in the limit of large sample size. In this paper, the 0.999 quantile of the annual distribution is used to quantify the impact of ignoring missing data for a broad range of frequency and severity distribution parameters and truncation levels.

Typically, the compound Poisson-Lognormal distribution is computed numerically by a Monte Carlo (MC) simulation method. Although Monte Carlo simulation is straightforward and robust, it has a severe limitation for the purposes of our study – the computing time required by MC is linearly proportional to the event frequency. This makes it very time consuming when the frequency is high. Alternatively, a few semi-analytical approaches for computing aggregate compound loss are applicable, for example: the recursive method of Panjer (1981) or direct inversion of the characteristic function for the compound distribution of Heckman and Meyers (1983) and Den Iseger (2006). These alternatives are analytically straightforward, but numerically difficult in terms of achieving high accuracy and computational efficiency *simultaneously*. Craddock, Heath and Platen (2000) gave an extensive survey of numerical techniques for inverting Laplace transforms and concluded that each of the many existing techniques has particular strengths and weaknesses, and no method works equally well for all classes of problems. For the study presented in this paper, we developed a numerical scheme specifically designed to invert the characteristic function of the compound Poisson-Lognormal distribution that ensures a high degree of accuracy in calculation of the 0.999 quantile. While a brief outline on the numerical steps will be provided in this paper, a complete description of the technical details of the algorithm is beyond the purpose of this paper and is the topic of a forthcoming paper.

Another common practice in calculating the annual loss distribution is to ignore the uncertainty in the fitted parameters. That is, the distribution conditional on the fitted parameters is used to estimate quantiles and final capital charge. Ignoring this uncertainty, which is always present in loss data modelling, may lead to a significant underestimation of capital charge. The uncertainty of parameter estimates can be treated using a Bayesian framework, see Shevchenko and Wüthrich (2006) for application of Bayesian inference in the operational risk context. In fact, when the variance of loss severity is large, a moderate 5% standard error in distribution parameters could lead to a substantial increase in the estimated capital charge. Ignoring truncation when fitting distributions (i.e. using "naive" or "shifted" models) usually leads to smaller fitting errors compared to the case of unbiased fitting. As a result, the final capital charge can be significantly underestimated if truncation is ignored in distribution fitting, as demonstrated by an example given in this paper. In addition to fitting errors (that will decrease as sample size increases), there are many other factors (for example, political, economical, legal, etc.) changing in time that do not allow precise knowledge of the severity and frequency distributions. One can model this by limiting standard errors of the parameters to some lower levels. This has been done in solvency approaches for the insurance industry, see e.g. Swiss Solvency Test in FOPI (2006), formulas (25)-(26).

The organisation of this paper is as follows. Section 2 describes the models studied in the paper. The method of characteristic functions used to calculate the compound Poisson-Lognormal distribution is briefly described in Section 3. Section 4 and Section 5 present results for the "shifted" and "naive" models respectively. Section 6 studies the impact of ignoring uncertainty in distribution parameters on the 0.999 quantile. Discussion of our results and conclusions is presented in Section 7.

## 2. Loss Data Model

The LDA for operational risk is based on modelling frequency and severity of losses, see e.g. Cruz (2002) and Frachot, Moudoulaud and Roncalli (2004). The annual loss in a risk cell under the LDA is

$$Z = \sum_{i=1}^{N} X_i, \tag{1}$$

where $N$ is the annual number of events (frequency) modelled as a random variable from a discrete distribution and $X_i, i = 1,...,N$ are the severities of the events modelled as independent and identically distributed random variables from a continuous distribution. It is assumed that severity and frequency of the events are independent. Note that independence assumed here is conditional on distribution parameters.

Hereafter we consider a single risk cell only and assume that $N$ and $X_i$ are modelled by the Poisson and Lognormal distributions respectively. The Lognormal density, $Lognormal(\mu,\sigma)$, is given by

$$f(x \mid \mu,\sigma) = \frac{1}{x\sqrt{2\pi\sigma^2}} \exp\left(-\frac{(\ln x - \mu)^2}{2\sigma^2}\right), \quad 0 < x < \infty, \tag{2}$$

and the Poisson density, $Poisson(\lambda)$, is

$$p(k \mid \lambda) = \frac{\lambda^k}{k!} \exp[-\lambda], \quad \lambda > 0, k = 0,1,... \quad . \tag{3}$$

Also, consider the density of a left-truncated Lognormal distribution (Frachot, Moudoulaud and Roncalli (2004))

$$f_L^{(T)}(x \mid \mu,\sigma) = \frac{f(x \mid \mu,\sigma)}{1 - F(L \mid \mu,\sigma)}; \quad L \leq x < \infty, \tag{4}$$

where $L$ is the truncation level and

$$F(L \mid \mu, \sigma) = \int_0^L f(x \mid \mu, \sigma) dx. \tag{5}$$

is the Lognormal cdf.

Assuming that losses originating from $f(x \mid \mu, \sigma)$ and $p(k \mid \lambda)$ are recorded above known reporting level $L$, the true parameters $\mu$ and $\sigma$ can be estimated by fitting the distribution $f_L^{(T)}(x \mid \mu, \sigma)$ to the truncated sample data using the method of maximum likelihood. Then the true intensity $\lambda$ can be estimated using the relationship (Frachot, Moudoulaud and Roncalli (2004))

$$\lambda = \frac{\theta}{1 - F(L \mid \mu, \sigma)}, \tag{6}$$

where $\theta$ is the intensity of the losses above $L$. The parameter estimates are asymptotically unbiased and converge to the true parameters as the sample size increases. Hereafter, the above model is called the "**unbiased model**".

Suppose that the shifted Lognormal density

$$f_L^{(S)}(x \mid \mu_s, \sigma_s) = \frac{1}{(x-L)\sqrt{2\pi\sigma_s^2}} \exp\left(-\frac{(\ln(x-L) - \mu_s)^2}{2\sigma_s^2}\right), \quad L \leq x < \infty, \tag{7}$$

is fitted to the truncated data using the method of maximum likelihood. In the limit of large sample size, the parameters of this distribution $\mu_S$ and $\sigma_S$ can be determined in terms of the true parameters $\mu$ and $\sigma$ as follows:

$$\mu_S = \int_L^\infty \ln(x-L) f_L^{(T)}(x \mid \mu, \sigma) dx, \tag{8}$$

$$\sigma_S^2 = \int_L^\infty [\ln(x-L)]^2 f_L^{(T)}(x \mid \mu, \sigma) dx - \mu_S^2. \tag{9}$$

Integrals in (8) and (9) can be efficiently calculated using standard integration routines (e.g. adaptive integration routine QDAGI from IMSL library). The above model will be referred to as the "**shifted model**". In this model the frequency is modelled by $Poisson(\theta)$, i.e. losses below $L$ are ignored.

The third model considered in this paper, referred as the "**naive model**", is based on the un-truncated Lognormal with density $f(x \mid \mu_u, \sigma_u)$ defined by (2) and fitted to data above the threshold $L$ using the method of maximum likelihood. Similar to the "shifted model", in the limit of large sample size, parameters $\mu_U$ and $\sigma_U$ can be determined via the true parameters $\mu$ and $\sigma$ as follows (see Chernobai *et al*. 2005):

$$\mu_U = \int_L^\infty \ln(x) f_L^{(T)}(x\,|\,\mu,\sigma)dx, \qquad (10)$$

$$\sigma_U^2 = \int_L^\infty (\ln x)^2 f_L^{(T)}(x\,|\,\mu,\sigma)dx - \mu_U^2. \qquad (11)$$

Unlike the "shifted model" the integrals in (10) and (11) for the "naive model" can be evaluated analytically. A detailed discussion will be given in Section 5. The frequency under the "naive model" is modelled by *Poisson*($\theta$), *i.e.* losses below the threshold are ignored when the intensity of loss events is estimated.

Both the "naive model" and the "shifted model" are biased for finite truncation. Their parameter estimates will never converge to the parameters of the true distribution as the sample size increases. Denote the annual loss under the "unbiased model" (i.e. based on the true distributions) as

$$Z^{(0)} = \sum_{i=1}^N X_i, \quad N \sim p(.\,|\,\lambda), \quad X_i \overset{iid}{\sim} f(.\,|\,\mu,\sigma) \qquad (12)$$

the annual loss under the "shifted model" as

$$Z^{(S)} = \sum_{i=1}^N X_i, \quad N \sim p(.\,|\,\theta), \quad X_i \overset{iid}{\sim} f_L^{(S)}(.\,|\,\mu_S,\sigma_S), \qquad (13)$$

and the annual loss under the "naive model" as

$$Z^{(U)} = \sum_{i=1}^N X_i, \quad N \sim p(.\,|\,\theta), \quad X_i \overset{iid}{\sim} f(.\,|\,\mu_U,\sigma_U). \qquad (14)$$

Here $f(.\,|\,\mu,\sigma)$ and $f(.\,|\,\mu_U,\sigma_U)$ are Lognormal density and $f_L^{(S)}(.\,|\,\mu_S,\sigma_S)$ is given by (7). Also, denote corresponding 0.999 quantiles as $Q^{(0)}$, $Q^{(S)}$ and $Q^{(U)}$, i.e.

$$\Pr(Z^{(0)} \le Q^{(0)}) = \Pr(Z^{(S)} \le Q^{(S)}) = \Pr(Z^{(U)} \le Q^{(U)}) = 0.999. \qquad (15)$$

The difference (bias) between $Q^{(0)}$ and $Q^{(s)}$, and between $Q^{(0)}$ and $Q^{(U)}$ will be studied in subsequent sections.

## 3. Computing the aggregate loss distribution

The most robust and straightforward numerical method for computing a compound distribution is Monte Carlo (MC) simulation. However, accurate computation of the 0.999 quantile requires a large number of MC simulations. The number of simulations $n$ should be increased by a factor of 100 to add one decimal place of precision, owing to the $O(1/\sqrt{n})$ convergence rate. In addition, the MC computational work is linearly proportional to event frequency. This makes it very time consuming when the frequency is high. The compound distribution can also be calculated using the method of characteristic functions (CF) as described below. The characteristic function of the severity density $f(x)$, $x \geq 0$ is

$$\varphi(t) = \int_{-\infty}^{\infty} f(x)e^{itx}dx, \qquad (16)$$

where $i = \sqrt{-1}$ is the standard imaginary unit. Then the characteristic function of the annual loss Z in model (1), with N distributed as $Poisson(\lambda)$, is

$$\chi(t) = \sum_{k=0}^{\infty}[\varphi(t)]^k \frac{e^{-\lambda}\lambda^k}{k!} = \exp[\lambda\varphi(t) - \lambda]. \qquad (17)$$

The density of Z can be found by inverting the above characteristic function. Then, given that Z is non-negative, the distribution of Z can be calculated as

$$H(z) = \frac{2}{\pi}\int_{0}^{\infty} \text{Re}[\chi(t)]\frac{\sin(tz)}{t}dt. \qquad (18)$$

Equation (18) will be used to calculate the annual loss distribution under the "unbiased", "shifted" and "naive" models.

Numerical computation of the characteristic function and its inversion to get the compound distribution involves one-dimensional integrals only, albeit over a semi-infinite line. The computer time required to calculate and invert the characteristic function may increase with Poisson parameter $\lambda$ as the integrand becomes more oscillatory. However, the increase in time is certainly not so significant as for the MC method. This makes the CF method a more efficient tool for computing the compound loss distribution at high event frequencies. Much work has been done over the last few decades in the area of numerical inversion of characteristic functions, for example Heckman and Meyers (1983) and Den Iseger (2006), to mention a few. Various issues should be addressed, such as singularity at the origin, treatment of long tails in the infinite integration, choices of quadrature rules, etc. We believe no single approach is superior to all others under all circumstances, as the extensive survey of Craddock, Heath and Platen (2000) has concluded. The specific distribution function and one's objectives dictate what method is most suitable. A tailor-made numerical algorithm for a specific distribution with a specific requirement on accuracy and efficiency is perhaps the best approach. A new numerical scheme for computing the Poisson-Lognormal compound distribution was devised specifically for this study, meeting the dual requirement of high accuracy and efficiency. A complete description of the technical details of the algorithm is beyond the purpose of this paper and is the topic of a forthcoming paper. Here we provide a brief outline on the key steps of our numerical algorithm for integrating (16) and (18):

- The integration of (16) for computing the characteristic function is done using the modified Clenshaw-Curtis integration method (Clenshaw and Curtis 1960; Piessens, Doncker-Kapenga, Überhuber and Kahaner 1983);
- The integration of (18) for the compound distribution is accomplished by an adaptive Gaussian quadrature taking into consideration the varying oscillation frequency and magnitude of the integrand;
- The accuracy and efficiency of integration of (18) is further improved by a special piecewise-linear approximation of the integrand tail.

It is worth noting that the Fast Fourier Transform (FFT) technique can also be used for efficient computing of the compound distribution (18).

## 4. Comparison of shifted and unbiased models

This section studies the bias introduced by the "shifted model" (7, 13) into the 0.999 quantile of the annual loss distribution. The bias is quantified by the relative difference

$$\delta \equiv \frac{Q^{(s)} - Q^{(0)}}{Q^{(0)}}, \qquad (19)$$

where $Q^{(0)}$ is the true value of the 0.999 quantile, i.e. model (12), and $Q^{(S)}$ is the quantile value under the "shifted model" (13). This relative difference is invariant under changes to the scale parameter $\mu$. Without loss of generality, we set $\mu = 3$ and consider only changes in $\sigma$, $\lambda$ and the truncation level $L$. Instead of using the absolute truncation value $L$, we present our results in terms of the percentage of points truncated $\Psi \equiv F(L\,|\,\mu,\sigma) \times 100\%$, where $F(L\,|\,\mu,\sigma)$ is the Lognormal cdf.

For each set of parameters the following calculations were performed:
1. Given $\mu, \sigma$ and truncation $\Psi$, calculate $\mu_s$ by (8) and $\sigma_s$ by (9). Given $\theta$ (the frequency of events above $L$) compute the true $\lambda$ by (6);
2. Calculate the true 0.999 quantile $Q^{(0)}$ of the annual loss distribution for the true model (12) with parameters $\mu, \sigma, \lambda$;
3. Calculate the 0.999 quantile $Q^{(s)}$ of the annual loss distribution under the "shifted model" (13) with parameters $\hat{\mu}_s$, $\hat{\sigma}_s$, $\theta$;
4. Calculate the relative difference $\delta \equiv (Q^{(s)} - Q^{(0)})/Q^{(0)}$.

The parameter ranges used in this study were:

$\sigma = (1.0,\ 2.0)$;
$\theta = (0.01, 0.1, 1.0, 10, 10^2, 10^3, 10^4, 10^5, 10^6)$;
$\Psi = (0\%, 1\%, 5\%, 10\%, 20\%, 30\%, 40\%, 50\%, 60\%, 70\%)$.

The ranges tested here were chosen to cover parameter values commonly encountered in operational risk modelling, see e.g. Moscadelli (2004). Some cases go beyond the realistic range, for example if $\theta$ is one million and truncation is 70% then the adjusted "true" $\lambda$ is over 3.33 million. In all tests, the quantiles were calculated by the CF method, represented by solid symbols in Figures (the numerical errors are always less than the symbol size). However, for small to moderate values of $\theta$ and $\lambda$ we also ran 1,000,000 Monte Carlo simulations in some selected cases for comparison purposes. The results from the MC method were in perfect agreement with the CF method and will not be shown here. To help the reader, the MC simulation procedure calculating the quantile $Q^{(0)}$ was accomplished as follows:

**Step1**. Simulate $N$ from the $Poisson(\lambda)$.

**Step2**. Simulate severities $X_i, i = 1,...,N$ from the $Lognormal(\mu, \sigma)$, and calculate $Z = \sum_{i=1}^{N} X_i$.

**Step3**. Repeat **Step1** to **Step2** $K$ times and identify possible realisations of annual loss $Z_i, i = 1,...,K$. Note that all random numbers generated in the above steps are independent.

**Step4**. Estimate the 0.999 quantile of the annual loss using the sample $Z_i, i = 1,...,K$.

The same procedure was used to calculate the quantile $Q^{(s)}$ by the MC method, except that the severities and frequencies were sampled from the shifted Lognormal distribution $f_L^{(S)}(.|\mu_s, \sigma_s)$ and $Poisson(\theta)$ respectively.

### 4.1. Results for $\sigma = 1$.

Figure 1 shows curves of $\delta$ as a function of the truncation percentage $\Psi$ for $\theta$ at 0.01, 0.1 and 1. Over-prediction of the 0.999 quantile by the "shifted model" increases when $\theta$ increases from 0.01 to 1. The largest over-prediction of approximately 100% is observed for truncations $\Psi > 10\%$ and $\theta = 1$. As shown in Figure 2, at higher frequency ($10^6 \geq \theta \geq 10$) the trend is reversed: over-prediction by the shifted model *decreases* with frequency, and actually changes to under-prediction at sufficiently high frequency and truncation levels. The curves for $\delta$ at $\theta = 10^5$ and $10^6$ are not shown in Figure 2 as they are almost indistinguishable from each other and are very close to the curve for $\theta = 10^4$. Both curves are hardly distinguishable from that predicted by central limit theory as discussed later.

It is interesting to examine the plots of $\delta$ as a function of the frequency parameter $\theta$. Figure 3 shows these curves at three truncation levels. Here, over-prediction of the "shifted model" increases with $\theta$ initially and then decreases until it becomes flat at very high frequencies. The influence of truncation is uniform at small frequency (the curves are close to each other), and the curves diverge at high frequencies, where over-prediction changes to under-prediction, as truncation level increases. This behaviour is as expected: at high frequencies and high truncation levels the shifted model omits too many loss events and, as a result, it underestimates aggregate loss. However, this underestimation is less than 2% at most for truncation levels up to $\Psi = 50\%$. From Figure 3 one can also see that $\delta$ approaches a constant for each given truncation level when the frequency is very high.

Another interesting point to note is the "sharp" onset of the bias for the "shifted model": at $\Psi = 5\%$ or even at $\Psi = 1\%$ all the curves in Figure 1 and Figure 2 show significant bias, despite the fact that $\delta$ approaches zero when $\Psi \to 0\%$. This can be explained as a high sensitivity of the 0.999 quantile to the severity parameters, especially the shape parameter. A small change in the value of severity parameter results in a substantial change in the 0.999 quantile.

### 4.2. Results for $\sigma = 2$.

As shown in Figure 4 and Figure 5, the main difference between cases at $\sigma = 2$ and those at $\sigma = 1$ is that the relative error $\delta$ becomes much smaller at $\sigma = 2$. Almost all results are within the $\pm 10\%$ bounds, irrespective of the loss frequency. This reflects the fact that with higher variance (fatter tail) in the loss data a shift on the left has relatively smaller impact on high quantiles, compared with a smaller variance. Another noticeable difference is that at $\sigma = 2$, the $\delta$ curves for $\theta \geq 10000$ are still easily distinguishable as shown in Figure 5, unlike the cases at $\sigma = 1$ where the $\delta$ curves for $\theta \geq 10000$ are hardly distinguishable. Again, similar to cases for $\sigma = 1$, the onset of bias is rather abrupt near $\Psi = 0\%$.

Curves of $\delta$ as a function of the frequency parameter $\theta$ for $\sigma = 2$ at three truncation levels are shown in Figure 6. For low truncation ($\Psi \leq 20\%$) the "shifted model" always over-predicts the 0.999 quantile. At moderate to high truncation the model over-predicts at low frequencies, but under-predicts as loss frequency increases. The magnitude of under-prediction is very small at very high frequencies – less than 5%. Figure 6 also shows that at $\theta = 10^5$ the curves are still not flat, indicating $\delta$ still changes with $\theta$ even at such a high frequency, unlike the cases for $\sigma = 1$ where the $\delta$ curves become flat at $\theta = 10^5$..

### 4.3. Central limit theory approximation

At very high frequencies, central limit theory is expected to provide a good approximation to the distribution of the annual loss $Z$ in model (1). The mean and variance of $Z$ are given by

$$\text{mean}(Z) \equiv \mu_Z = E[N] \cdot E[X], \tag{20}$$

$$\text{var}(Z) \equiv \sigma_Z^2 = E[N] \cdot \text{var}(X) + \text{var}(N) \cdot (E[X])^2, \tag{21}$$

where $E[X] = E[X_1] = ... = E[X_N]$, $\text{var}(X) = \text{var}(X_1) = ... = \text{var}(X_N)$. If $N$ is distributed from $Poisson(\lambda)$ and $X_i, i = 1,...,N$ are iid from $Lognormal(\mu,\sigma)$ then $\mu_Z = \lambda \exp(\mu + 0.5\sigma^2)$ and $\sigma_Z^2 = \lambda \exp(2\mu + 2\sigma^2)$. According to central limit theory, the Poisson-Lognormal compound distribution at very high frequency should approach the Normal distribution with mean $\mu_Z$ and standard deviation $\sigma_Z$. In Figure 7, the $\delta$ curves calculated using central limit theory approximation with $\theta = 10^6$ and $\sigma = 1.0, 2.0$ are compared with the numerical CF method results (solid symbols) using (18). Indeed, at very high frequencies, the numerically calculated 0.999 quantile of the compound distribution is well approximated by the 0.999 quantile of the Normal distribution predicted by central limit theory. The sudden onset of model bias near $\Psi = 0$ is again apparent in Figure 7.

## 5. Comparison of the "naive" and "unbiased" models

We have demonstrated that although the "shifted model" always underestimates the loss event frequency, it often overestimates the 0.999 quantile of the annual loss distribution due to overestimation of loss severity. In this section we analyse the bias introduced by the "naive model" (10,11,14) and show that for all practical cases it underestimates the 0.999 quantile of annual loss.

### 5.1. Underestimation of loss severity by the "naive model"

In the analysis by Chernobai *et al.* (2005), it was shown that the "naive model" underestimates the shape parameter $\sigma$ under the condition $\ln L < \mu$ and always overestimates the scale parameter $\mu$. It is not obvious whether the "naive model" will over-predict or under-predict the quantiles of severity. Chernobai *et al.* (2005) gave a few numerical examples at 5% and 10% truncation to demonstrate that the capital charge quantile of the compound Poisson-Lognormal distribution is underestimated by the "naive model", partly due to its underestimation of event frequency. Below we prove that the "naive model" will always underestimate the shape parameter $\sigma$, i.e. the condition $\ln L < \mu$ is not required. Also, we derive the condition for the "naive model" to underestimate the high quantiles of loss severity.

Substituting (4) into (10) and (11) we obtain the following closed form solutions for $\mu_U$ and $\sigma_U$:

$$\mu_U = \frac{\sigma}{[1-F(L|\mu,\sigma)]\sqrt{2\pi}} \exp\left[-\frac{(\ln L - \mu)^2}{2\sigma^2}\right] + \mu, \qquad (22)$$

and

$$\sigma_U^2 = \frac{\sigma}{[1-F(L|\mu,\sigma)]\sqrt{2\pi}} \exp\left[-\frac{(\ln L - \mu)^2}{2\sigma^2}\right](\ln L + \mu) + \sigma^2 + \mu^2 - \mu_U^2. \qquad (23)$$

Equation (22) and (23) are identical to those derived by Chernobai *et al.* (2005).

Define $t = (\ln L - \mu)/\sigma$, then the above expressions can be simplified to

$$\mu_U = \frac{f_N(t)\sigma}{1-F_N(t)} + \mu = A(t)\sigma + \mu, \qquad (24)$$

and

$$\sigma_U^2 = (\mu_U - \mu)(\ln L + \mu) + \sigma^2 + \mu^2 - \mu_U^2 = (\mu_U - \mu)(\ln L - \mu_U) + \sigma^2, \qquad (25)$$

where $f_N(t)$ and $F_N(t)$ are the pdf and cdf respectively for the standard Normal distribution, and $A(t) = f_N(t)/(1-F_N(t))$. Note that, $F_N(t) = F(L|\mu,\sigma)$ is the fraction of truncated losses. As $A(t) > 0$ is always true, then formula (24) implies that $\mu_U - \mu > 0$. Thus, according to (25), $\sigma_U < \sigma$ if and only if $\ln L < \mu_U$. From (24) and $t = (\ln L - \mu)/\sigma$, it is easy to show that $\ln L < \mu_U$ is equivalent to $t < A(t)$ or

$$B(t) = A(t) - t = \frac{f_N(t)}{1 - F_N(t)} - t > 0. \tag{26}$$

Thus the "naive model" underestimates the severity variance if and only if $B(t) > 0$, which is a property of the standard Normal distribution alone. Figure 8 shows plot of the function $B(t)$ in the range $(-6 \leq t \leq 6)$. Since $F_N(t)$ is also a fraction of truncated loss and is a monotonically increasing function of $t$, we can also plot $B(t)$ as a function of $F_N(t)$. The range $(-6 \leq t \leq 6)$ in Figure 8 covers all possible truncations from virtually 0% to 100%. It can be shown that $B(t) \to 0$ as $t \to \infty$ as follows:

$$\lim_{t \to \infty} B(t) = \lim_{t \to \infty} \left\{ \frac{f_N(t)}{1 - F_N(t)} - t \right\} = -\frac{df_N(t)/dt}{dF_N(t)/dt} - t = \frac{f_N(t) \times t}{f_N(t)} - t = 0. \tag{27}$$

This demonstrates that the "naive model" always underestimates the shape parameter $\sigma$ of the Lognormal distribution. This statement is completely determined by a property of the standard Normal distribution as shown in the necessary and sufficient condition (26).

Now let us derive the condition for underestimation of high quantiles of the true Lognormal distribution. Denote $q_\alpha$ as the quantile of the standard Normal distribution at the confidence level $\alpha$, i.e. $F_N(t = q_\alpha) = \alpha$, then $Q_\alpha^{(LN)} = \exp(\mu + \sigma q_\alpha)$ is the quantile of the true Lognormal distribution at this level. Thus the condition that the "naive model" underestimates $Q_\alpha^{(LN)}$ is

$$q_\alpha \sigma_U + \mu_U < q_\alpha \sigma + \mu, \tag{28}$$

which can be re-arranged as

$$1 - \frac{A(t)}{q_\alpha} > \frac{\sigma_U}{\sigma}. \tag{29}$$

Using (24) and (25) we find that

$$\frac{\sigma_U}{\sigma} = \sqrt{1 + A(t)[t - A(t)]}. \tag{30}$$

Thus the necessary and sufficient condition for the "naive model" to underestimate $Q_\alpha^{(LN)}$ is

$$C(t, \alpha) \equiv \frac{A(t)}{q_\alpha} + \sqrt{1 + A(t)[t - A(t)]} - 1 < 0. \tag{31}$$

Again, this condition is completely determined by a property of the standard Normal distribution alone, for any given confidence level. One can easily see that the above inequality does not hold for sufficiently small $q_\alpha$, or equivalently, a sufficiently small $\alpha$. Figure 9 shows plots of $C(t,\alpha)$ as a function of $F_N(t)$. Clearly, the "naive model" underestimates $Q_{0.99}^{(LN)}$ when truncation is below 66% and underestimates $Q_{0.999}^{(LN)}$ when truncation is below 92%. Note that it is not practical to fit a distribution if more than 60% of data points are truncated.

### 5.2. Underestimation of the 0.999 quantile by the "naive model"

Since the "naive model" underestimates loss frequency and high quantiles of severity, it is expected that it will underestimate the 0.999 quantile of the compound Poisson-Lognormal distribution. Figures 10-13 demonstrate this underestimation for a broad range of parameters. Also, we have performed calculations for many other parameter values not shown here (including very small $\sigma$) and concluded that the "naive model" underestimates the 0.999 quantile of the annual loss for all practical (and all reasonable) cases. The bias $\delta$ presented in Figures 10-13 is defined as $(Q^{(U)} - Q^{(0)})/Q^{(0)}$, where $Q^{(U)}$ and $Q^{(0)}$ are the values of 0.999 quantile of the annual loss under the "naive model" (14) and "unbiased model" (12) respectively.

Figures 10 and 11 show the bias when $\sigma = 1$ and $\theta$ ranging from 0.01 to 1000. The magnitude of underestimation appears to be monotonically increasing up to approximately 35% as truncation level increases. Figures 12 and 13 show results for the bias when $\sigma = 2$. Unlike the "shifted model" which has smaller bias at higher severity variance, the "naive model" underestimates 0.999 quantile even more severely at higher values of $\sigma$. The magnitudes of underestimation at low frequency are all larger than those at $\sigma = 1$ for the same frequency. The magnitudes of underestimation at frequencies $\theta \geq 10$ are more than 50% for $\sigma = 2$. Again, the magnitude of underestimation $\delta$ initially increases with frequency when $\theta \leq 1$ and then decreases with frequency when $\theta \geq 10$.

## 6. Impact of parameter uncertainty

The biases introduced by the "naive" and "shifted" models, studied in the above sections, are the biases in the limit of large sample size. The parameters fitted using real data are estimates that have statistical fitting errors due to finite sample size. The true parameters are not known. In our experience with banks, typically, uncertainty in fitted parameters is ignored when capital is quantified. That is parameters are fixed to their point estimates (e.g. maximum likelihood estimates) when the annual loss distribution and its 0.999 quantile are calculated. The impact of parameter uncertainty on quantile estimates can be taken into account using a Bayesian framework, see Shevchenko and Wüthrich (2006) for an application of Bayesian framework to operational risk. Consider model (1) where frequency and severity are modelled by the densities $p(.|\boldsymbol{\alpha})$ and $f(.|\boldsymbol{\beta})$ respectively. Here, $\boldsymbol{\alpha}$ and $\boldsymbol{\beta}$ are the distribution parameter vectors for frequency and severity respectively. Denote the vector of all parameters by $\boldsymbol{\gamma} = (\boldsymbol{\alpha}, \boldsymbol{\beta})$ and the random vector of observations (severities and frequencies) by $\mathbf{Y}$. Given $\boldsymbol{\gamma}$, denote the density of annual loss (1) as $g(z|\boldsymbol{\gamma})$. Then the density of the predictive distribution for the next year is

$$h(z \mid \mathbf{Y}) = \int g(z \mid \boldsymbol{\gamma}) \hat{\pi}(\boldsymbol{\gamma} \mid \mathbf{Y}) d\boldsymbol{\gamma}, \qquad (32)$$

where $\hat{\pi}(\boldsymbol{\gamma} \mid \mathbf{Y})$ is the joint posterior density of the parameters given observations $\mathbf{Y}$. From Bayes' rule

$$\hat{\pi}(\boldsymbol{\gamma} \mid \mathbf{Y}) \propto l(\mathbf{Y} \mid \boldsymbol{\gamma}) \pi(\boldsymbol{\gamma}), \qquad (33)$$

where $l(\mathbf{Y} \mid \boldsymbol{\gamma})$ is the likelihood of observations and $\pi(\boldsymbol{\gamma})$ is a prior distribution for the parameters (a prior distribution can be specified by an expert or fitted using external data). The quantile $\hat{\tilde{Q}}$ of $h(z \mid \mathbf{Y})$ can be easily calculated, for example, by the following MC procedure:

**Step 1**. Simulate $\boldsymbol{\alpha}$ and $\boldsymbol{\beta}$ from their joint distribution $\hat{\pi}(\boldsymbol{\gamma} \mid \mathbf{Y})$.

**Step 2**. Given $\boldsymbol{\alpha}$ and $\boldsymbol{\beta}$ calculate annual loss $Z = \sum_{i=1}^{N} X_i$ by simulating $N$ from the frequency distribution $p(. \mid \boldsymbol{\alpha})$ and $X_i, i = 1,...,N$ from the loss severity distribution $f(. \mid \boldsymbol{\beta})$.

**Step 3**. Repeat **Step 1** to **Step 2** $K$ times to identify possible realisations of total annual loss $Z_j, j = 1,....,K$.

**Step 4**. Estimate the 0.999 quantile, $\hat{\tilde{Q}}$, of the total annual loss using simulated sample $Z_j, j = 1,....,K$.

In the above we model both the process uncertainty (severity and frequencies are random variables) and the parameter uncertainty (parameters are simulated from their posterior distribution). The parameter uncertainty comes from the fact that we do not know the true values of the parameters. In this framework, the capital charge should be based on the 0.999 quantile $\hat{\tilde{Q}}$ of the predictive distribution.

Consider loss events above threshold $L$ over a period of $M$ years, i.e. with annual frequencies $N_m, m = 1,...,M$ and severities $X_j, j = 1,...,J$ ($J = \sum_{m=1}^{M} N_m$). In the case of "unbiased model" considered in this paper, i.e. fitting truncated Lognormal distribution (4), the parameters are $\boldsymbol{\gamma} = (\mu, \sigma, \lambda)$ and the likelihood in (33) can be written as

$$l(\mathbf{Y} \mid \boldsymbol{\gamma}) = \prod_{j=1}^{J} f_L^{(T)}(X_j \mid \mu, \sigma) \times \prod_{m=1}^{M} p(N_m \mid \lambda \times [1 - F(L \mid \mu, \sigma)]), \qquad (34)$$

where $f_L^{(T)}(.)$ is the truncated Lognormal density (4) and $p(.)$ is the Poisson density (3). In the case of the "naive model" and the "shifted" model the parameters are $\boldsymbol{\gamma} = (\mu_S, \sigma_S, \theta)$ and $\boldsymbol{\gamma} = (\mu_U, \sigma_U, \theta)$ respectively. The corresponding likelihoods are straightforward.

Sometimes it is possible to find the posterior distribution $\hat{\pi}(\boldsymbol{\gamma} \mid \mathbf{Y})$ of the parameters in closed form. However, in general, $\hat{\pi}(\boldsymbol{\gamma} \mid \mathbf{Y})$ should be estimated numerically (e.g. using Markov Chain Monte Carlo methods described by Peters and Sisson (2006) in the context of operational risk). The mode of the posterior distribution $\hat{\boldsymbol{\gamma}} = \text{mode}(\boldsymbol{\gamma})$ can be used as a point estimator for the parameters. For large sample size (and continuous prior distribution), it is common to approximate $\ln \hat{\pi}(\gamma | \mathbf{Y})$ by a second-order Taylor series expansion around $\hat{\gamma}$. Then $\hat{\pi}(\gamma | \mathbf{Y})$ is approximately the multivariate Normal distribution with mean $\hat{\gamma}$ and covariance matrix $\hat{\mathbf{C}}$, which is the inverse of the matrix $\mathbf{I}$ whose elements are

$$\mathbf{I}_{ij} = -\frac{\partial^2 \ln \hat{\pi}(\gamma | \mathbf{Y})}{\partial \gamma_i \partial \gamma_j}\bigg|_{\gamma = \hat{\gamma}}. \quad (35)$$

The standard deviations and correlations of the parameters are $\hat{\tau}_i = \sqrt{\hat{\mathbf{C}}_{ii}}$ and $\hat{\rho}_{ij} = \hat{\mathbf{C}}_{ij} / \sqrt{\hat{\mathbf{C}}_{ii}\hat{\mathbf{C}}_{jj}}$ respectively. Under the standard fitting procedure, when prior knowledge is ignored (i.e. the case of so-called non-informative improper priors), $\hat{\pi}(\gamma | \mathbf{Y}) \propto l(\mathbf{Y} | \gamma)$ and the mode of the posterior distribution, $\hat{\gamma}$, is the same as the maximum likelihood estimator. Hereafter we assume non-informative priors.

Denote the estimate of the 0.999 quantile of the annual loss distribution $g(z|\hat{\gamma})$ as $\hat{Q}$ (calculated by e.g. MC or CF methods as described in previous sections). The impact of ignoring parameter uncertainty can be quantified by $\Delta \equiv (\hat{\hat{Q}} - \hat{Q})/\hat{Q}$. According to the above, denote the quantile estimates under the "unbiased", "shifted" and "naive" models as $\hat{Q}^{(0)}$, $\hat{Q}^{(S)}$ and $\hat{Q}^{(U)}$ respectively, when parameter uncertainty is ignored. Also, denote the quantile estimates under the "unbiased", "shifted" and "naive" models as $\hat{\hat{Q}}^{(0)}$, $\hat{\hat{Q}}^{(S)}$ and $\hat{\hat{Q}}^{(U)}$ correspondingly when parameter uncertainty is taken into account. We expect $\hat{\hat{Q}}$ to be larger then $\hat{Q}$, i.e. extra capital is required to cover extra uncertainty (which can be regarded as an extra risk) due to the fact that we do not know the true values of the distribution parameters. As the number of observations increases, $\hat{\hat{Q}}$ will converge to $\hat{Q}$ (it is also true if the prior distribution is continuous and does not vanish at the maximum of the likelihood).

The problem with the use of the simplified models that ignore data truncation, such as "naive" and "shifted" models, is not just the introduced bias but underestimation of extra capital required to cover parameter uncertainty. Typically these simplified models lead to smaller fitting errors. It is not difficult to find a realistic example where: $\hat{Q}^{(S)} > \hat{Q}^{(0)}$ but $\hat{\hat{Q}}^{(S)} < \hat{\hat{Q}}^{(0)}$ (i.e. "shifted model" overestimating the quantile leads to under-estimation when parameter uncertainty is taken into account). We simulate losses over a four year period from $Lognormal(\mu = 3, \sigma = 2)$, $Poisson(\lambda = 20)$, and truncate at a threshold level $L$ such that $\Psi = 10\%$, i.e. $L = F^{-1}(\Psi/100 | \mu, \sigma) \approx 1.548$, where $F^{-1}(.)$ is the inverse Lognormal cdf. In our example we obtained 62 losses after truncation. Then we fit three models as follows:

- **"unbiased model"**: fit the model (12) to the truncated data by maximizing complete likelihood (33) to obtain point estimates $\hat{\mu}$, $\hat{\sigma}$, $\hat{\lambda}$ of the parameters $\mu$, $\sigma$, $\lambda$. The standard deviations $\hat{\tau}_\mu$, $\hat{\tau}_\sigma$, $\hat{\tau}_\lambda$ and correlations $\hat{\rho}_{\mu\sigma}, \hat{\rho}_{\lambda\mu}, \hat{\rho}_{\lambda\sigma}$ of the parameters were calculated using (35).

- "**shifted model**": fit the model (13) to the truncated data using the maximum likelihood method to obtain the distribution parameter estimates $\hat{\mu}_S$, $\hat{\sigma}_S$, $\hat{\theta}$. The standard deviations $\hat{\tau}_\mu^{(S)}, \hat{\tau}_\sigma^{(S)}$ and $\hat{\tau}_\theta^{(S)}$ of the parameters $\mu_S, \sigma_S$ and $\theta$ respectively were calculated using (35). Correlations between parameters are zero in this case.
- "**naive model**": fit the model (14) to the truncated data using the maximum likelihood method to obtain the point estimates $\hat{\mu}_U, \hat{\sigma}_U$, $\hat{\theta}$ and standard deviations $\hat{\tau}_\mu^{(U)}, \hat{\tau}_\sigma^{(U)}, \hat{\tau}_\theta^{(U)}$ of the parameters $\mu_U, \sigma_U, \theta$ correspondingly. The standard deviations were calculated using (35). The correlations between parameters are zero in this case.

For each of the fitted model, the annual loss 0.999 quantiles $\hat{Q}$ and $\hat{\hat{Q}}$ (ignoring and taking parameter uncertainty into account respectively) were calculated using the Monte Carlo procedures described previously with $10^6$ simulations (numerical standard error was of the order of 1%). The posterior distribution $\hat{\pi}(\gamma | \mathbf{Y})$ used to simulate parameters was approximated by the multivariate Normal distribution with the mean e $\hat{\gamma}$ (maximum likelihood point estimator) and covariance matrix $\hat{\mathbf{C}}$ given by (35). Actually, for the "shifted" and "naive" models, the posterior distribution can be found in closed form (when non-informative priors are used). However, to be consistent with the "unbiased model" we used the Normal distribution approximation (in the case of parameter values used in this example, this approximation should not introduce material difference).

Table 1 lists the fitting results and quantile estimates. In this example, accounting for parameter uncertainty increases estimates of 0.999 quantile for all models, i.e. $\hat{\hat{Q}}^{(0)} > \hat{Q}^{(0)}$, $\hat{\hat{Q}}^{(S)} > \hat{Q}^{(S)}$, $\hat{\hat{Q}}^{(U)} > \hat{Q}^{(U)}$. This result was typical of results observed in many other test cases not shown here. That is, extra capital is required to cover parameter uncertainty risk. Also, as expected from the analysis in the previous sections $\hat{Q}^{(U)} < \hat{Q}^{(0)} < \hat{Q}^{(S)}$. That is, if parameter uncertainty is ignored, the "shifted model" over predicts the quantile and the "naive model" significantly underestimates the quantile when compared to the "unbiased model". In addition, the results show that $\hat{\hat{Q}}^{(U)} < \hat{\hat{Q}}^{(S)} < \hat{\hat{Q}}^{(0)}$. That is, both the "shifted" and "naive" models underestimate capital charge when the parameter uncertainty is taken into account. This is because the "shifted" and "naive" models lead to smaller fitting errors in comparison to the "unbiased model". Consequently, the change $\hat{\hat{Q}} - \hat{Q}$ for the "unbiased model" is the largest. Again, this is a typical result observed in many other test cases not presented here. Of course, as the number of observations increases, the impact of parameter uncertainty diminishes. However, for modest fitting errors 5-10% (often, in modelling operational risk data, the errors are larger) the impact of parameter uncertainty is significant. In the above example, we ignored the prior distributions that can be estimated by experts or using external data, see Shevchenko and Wüthrich (2006), Lambrigger *et al*. (2007). The use of the informative priors can significantly improve all estimators and allows for natural combining of the internal observations with expert opinion and external data.

# 7. Conclusions

In this paper, we undertook a systematic study of the impact of ignoring data truncation on estimation of operational risk under the "shifted" and "naive" models when loss frequency and severity are modelled by the Poisson and Lognormal distributions respectively. This study covers a wide range of parameters and truncation levels relevant to operational risk modelling. The bias introduced by these simplified models into the 0.999 quantile (in the limit of large sample size) was used to quantify the impact. The parameters of the "shifted" and "naive" models were calculated exactly via the parameters of the true Poisson and Lognormal distributions.

The "shifted model" underestimates loss event frequency and over-predicts loss severity. It was found that the impact can be considerable for small to moderate variance of the severity distribution, and it is relatively small when the variance is large. This impact initially increases with the Poisson parameter and then decreases. For moderate to high frequencies, the impact decreases with truncation level. The maximum observed bias introduced by the "shifted model" is about 100% over-prediction at low severity variance and moderate event frequency. Typically, "shifted model" leads to overestimation. However, at large truncation levels it may result in a small underestimation.

This study has demonstrated that the "naive model" underestimates both frequency and high quantiles of the severity. This leads to the underestimation of the annual loss 0.999 quantile for all practical cases. The underestimation can be as significant as 100%. The magnitude of the underestimation increases with the variance of loss severity, unlike the "shifted model", where the magnitude of the bias decreases with the variance of loss severity.

In our systematic study of the bias induced by the "shifted" and "naive" models we have ignored fitting errors (i.e. in the limit of large sample size). Often, in practice, the capital estimation is based on point estimators for distribution parameters (e.g. maximum likelihood estimators). In this case the "shifted model" typically overestimates the capital charge and its use can be justified as a conservative estimate. The "naive model" underestimates capital charge and would not be acceptable for capital calculations from a regulator's perspective. However, more accurate capital estimation should take into account parameter uncertainty, due to the fact that we do not know the true parameter values. Then the use of the "shifted model" is quite dangerous. An example was given to show that if parameter fitting error is taken into account, the "shifted model" (as well as "naive model") may significantly underestimate the capital charge when compared to the proper "unbiased model". This is mainly due to the fact that the "unbiased model" has larger fitting errors in the parameter estimates. Often operational risk loss data are limited and as a result fitting errors are large. In this case the extra capital required to cover the risk due to parameter uncertainty can be significant and cannot be ignored. Even for modest errors such as 5-10% the impact of parameter uncertainty on the 0.999 quantile is pronounced. In summary, a proper "unbiased model" (with parameter uncertainty taken into account) should be used to account for data truncation, otherwise the capital charge for operational risk can be substantially underestimated.

## Acknowledgement

We would like to thank Dr. Richard Jarrett and the anonymous referee for many constructive comments which have led to improvements in the manuscript.

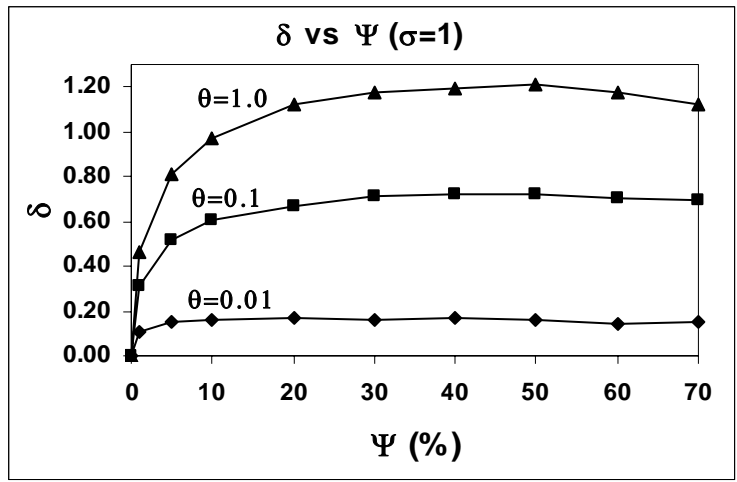

**Figure 1.** "Shifted model". The bias of the 0.999 quantile $\delta$ as a function of truncation $\Psi$ for small loss frequencies at $\sigma = 1$.

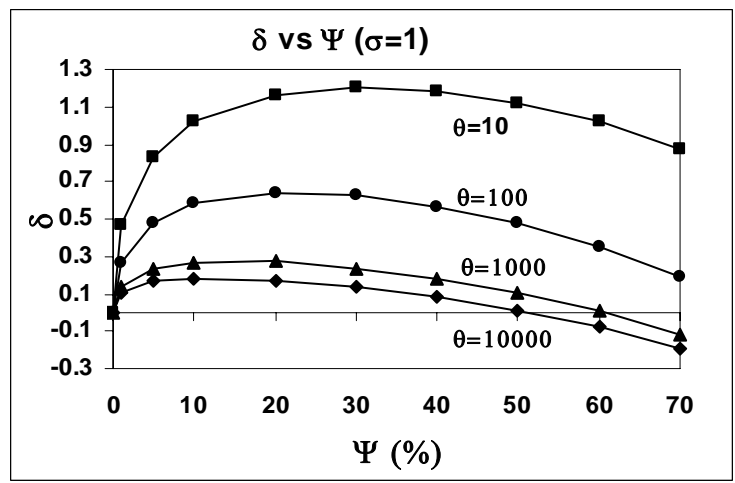

**Figure 2.** "Shifted model". The bias of the 0.999 quantile $\delta$ as a function of truncation $\Psi$ for medium and large loss frequencies at $\sigma = 1$.

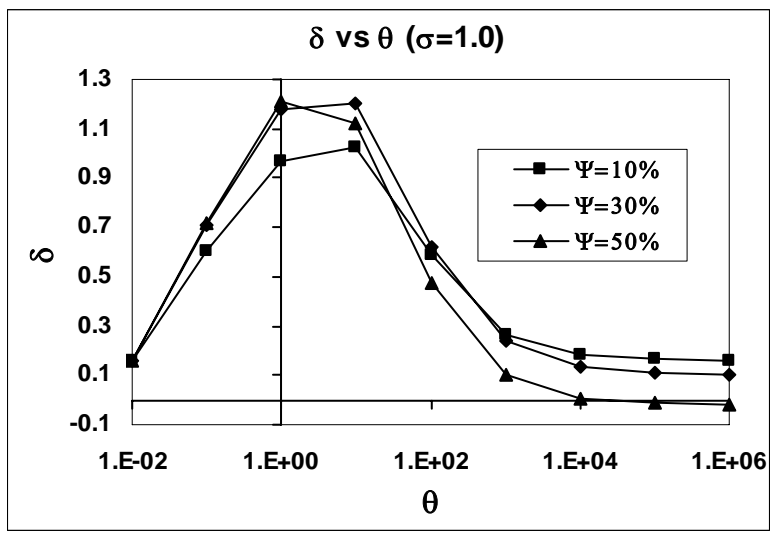

**Figure 3.** "Shifted model". The bias of the 0.999 quantile $\delta$ as a function of loss frequency for three truncation levels at $\sigma = 1$.

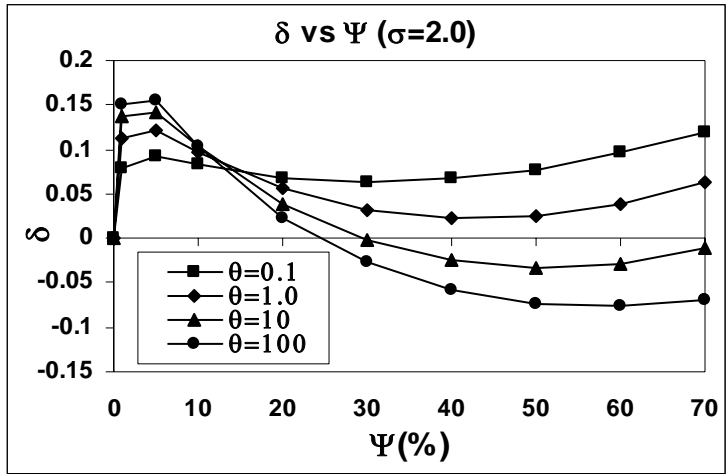

**Figure 4.** "Shifted model". The bias of the 0.999 quantile $\delta$ as a function of truncation $\Psi$ for small to medium loss frequencies at $\sigma = 2$.

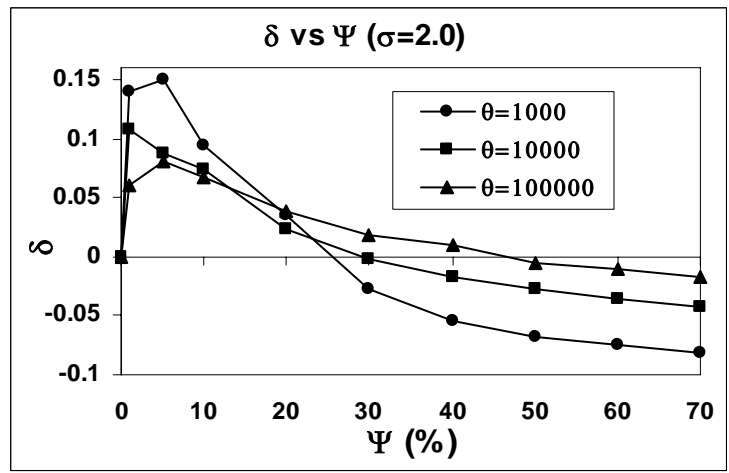

**Figure 5.** "Shifted model". The bias of the 0.999 quantile $\delta$ as a function of truncation $\Psi$ for large loss frequencies at $\sigma = 2$.

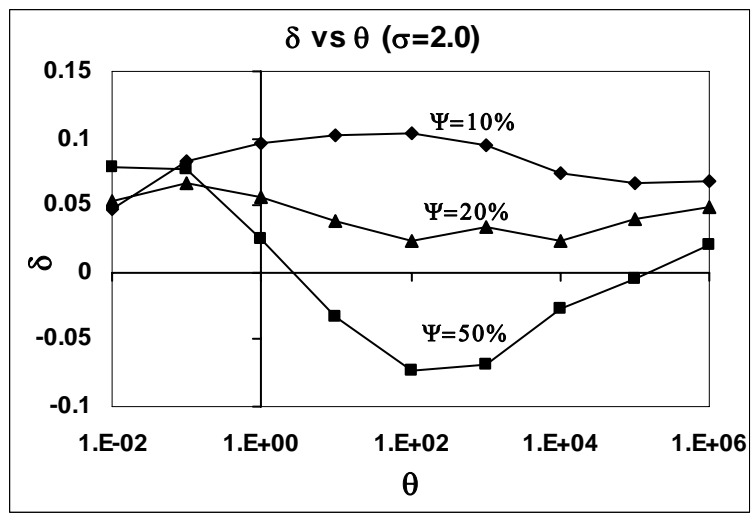

**Figure 6.** "Shifted model". The bias of the 0.999 quantile $\delta$ as a function of loss frequency for three truncation levels at $\sigma = 2$.

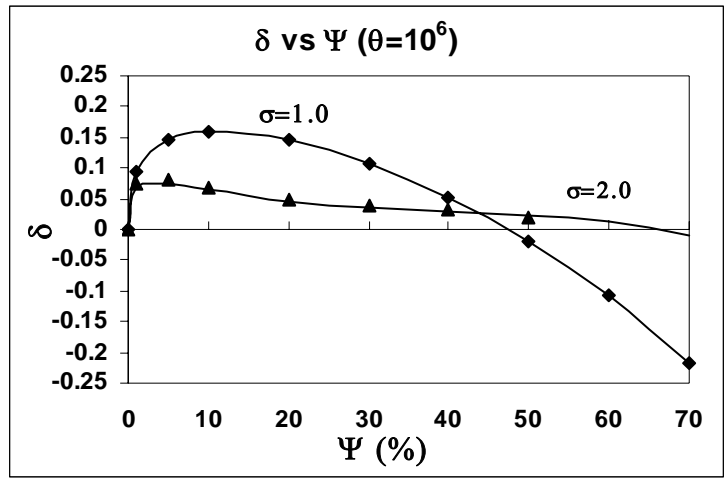

**Figure 7. "Shifted model". The bias of the 0.999 quantile $\delta$ as a function of truncation $\Psi$ at $\theta = 10^6$. Solid lines: results predicted by central limit theory; solid symbols: results via the CF method.**

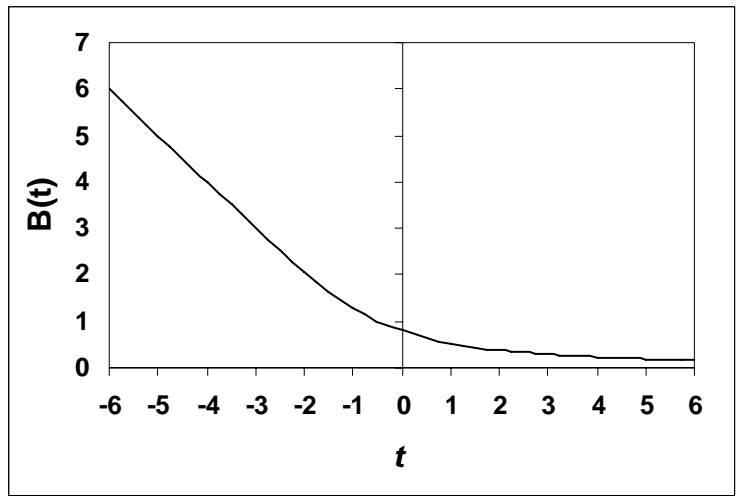

**Figure 8.** "Naive model". $B(t)$ as a function of $t = (\ln L - \mu)/\sigma$ to demonstrate that shape parameter of the Lognormal distribution is always underestimated by the "naive model".

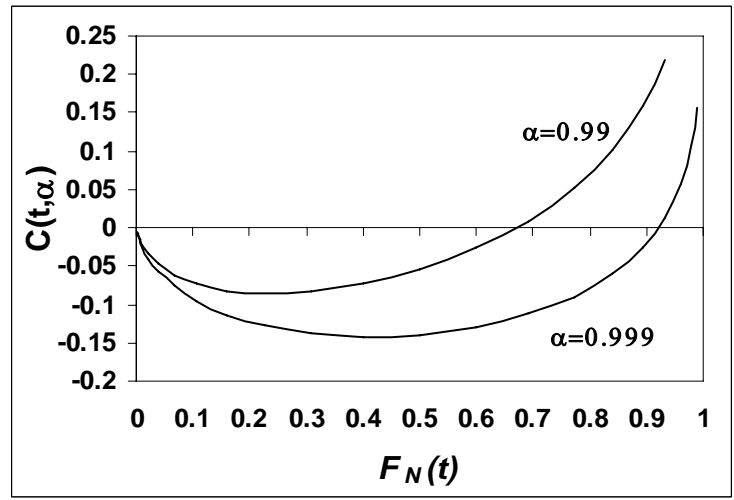

**Figure 9.** "Naive model". $C(t,\alpha)$ as a function of $F_N(t)$. Negative $C(t,\alpha)$ corresponds to under-estimation of quantile at level $\alpha$, $C(t,\alpha) > 0$ otherwise.

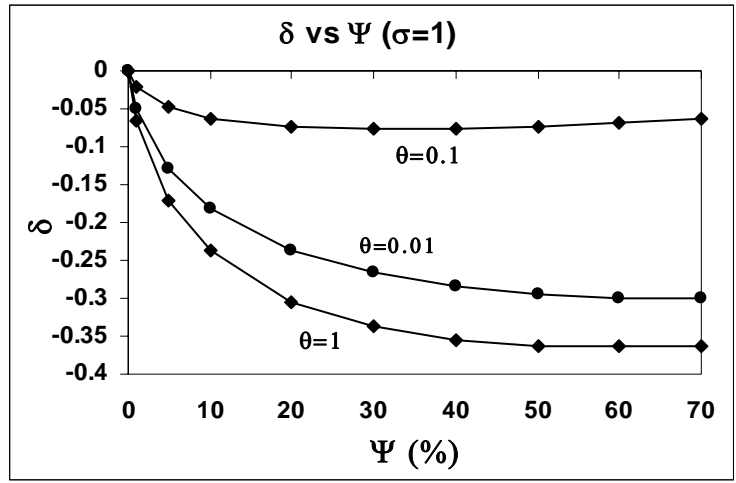

**Figure 10.** "Naive model". The bias of the 0.999 quantile $\delta$ as a function of truncation $\Psi$ for three loss frequencies (0.01, 0.1 and 1.0) at $\sigma = 1$.

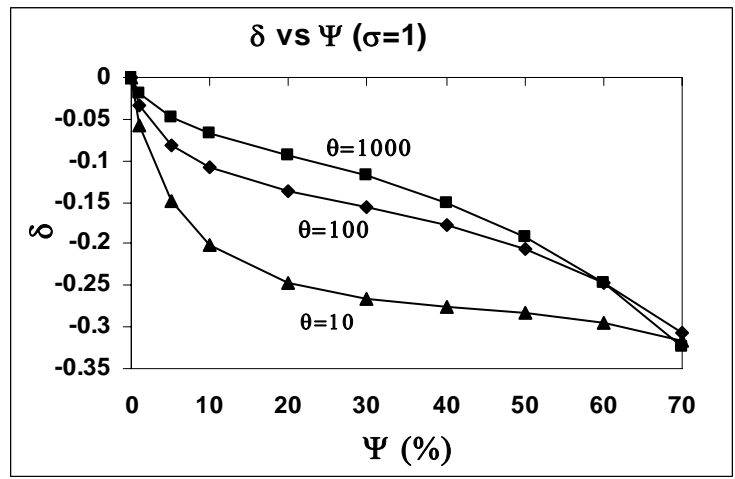

**Figure 11.** "Naive model". The bias of the 0.999 quantile $\delta$ as a function of truncation $\Psi$ for three loss frequencies (10, 100 and 1000) at $\sigma = 1$.

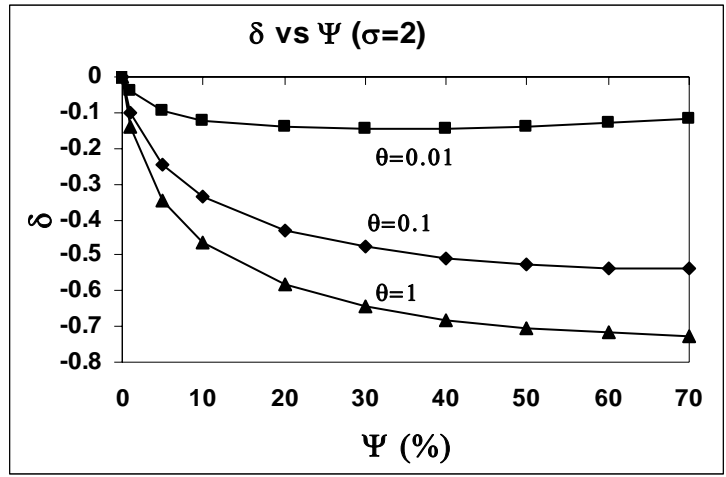

**Figure 12.** "Naive model". The bias of the 0.999 quantile $\delta$ as a function of truncation $\Psi$ for three loss frequencies (0.01, 0.1 and 1.0) at $\sigma = 2$.

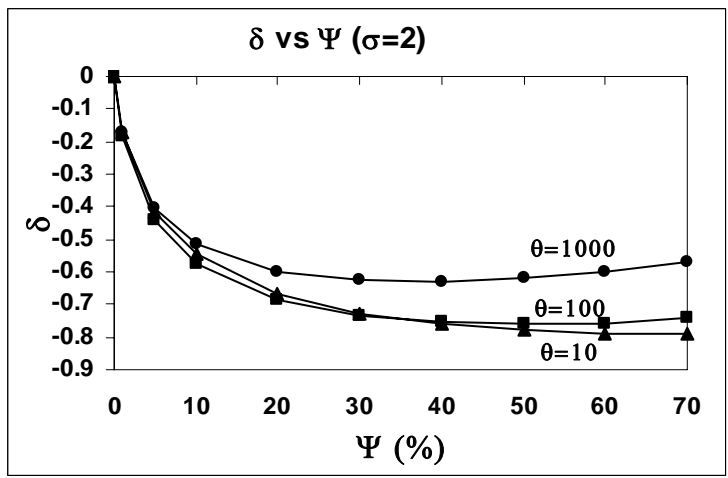

**Figure 13.** "Naive model". The bias of the 0.999 quantile $\delta$ as a function of truncation $\Psi$ for three loss frequencies (10, 100 and 1000) at $\sigma = 2$.

Table 1. Fitting results for the "unbiased", "shifted" and "naive" models to losses simulated from *Lognormal*(3,2) and *Poisson*(20) over a four year period and truncated at *L*=1.548 (i.e. $\Psi = 10\%$). The total number of simulated losses after truncation was 62. $\hat{Q}$ and $\hat{\hat{Q}}$ are the annual loss 0.999 quantiles calculated with parameter uncertainty ignored and taken into account respectively.

| | | |
|---|---|---|
| "Unbiased model" | $\hat{\mu} \approx 2.636, \quad \hat{\sigma} \approx 1.835, \quad \hat{\lambda} \approx 17.52$ <br> $\hat{\tau}_\mu \approx 0.255, \quad \hat{\tau}_\sigma \approx 0.232, \quad \hat{\tau}_\lambda \approx 2.415$ <br> $\hat{\rho}_{\mu\sigma} \approx -0.404, \hat{\rho}_{\mu\lambda} \approx -0.320, \hat{\rho}_{\sigma\lambda} \approx 0.331$ | $\hat{Q}_{0.999}^{(0)} \approx 1.8 \times 10^4$ <br> $\hat{\hat{Q}}_{0.999}^{(0)} \approx 5.1 \times 10^4$ <br> $\Delta \approx 1.8$ |
| "Shifted model" | $\hat{\mu}_S \approx 2.771, \quad \hat{\sigma}_S \approx 1.862, \quad \hat{\theta} = 15.5$ <br> $\hat{\tau}_\mu^{(S)} \approx 0.236, \quad \hat{\tau}_\sigma^{(S)} \approx 0.167, \quad \hat{\tau}_\theta^{(S)} \approx 1.969$ <br> $\hat{\rho}_{\mu\sigma}^{(S)} = \hat{\rho}_{\mu\theta}^{(S)} = \hat{\rho}_{\sigma\theta}^{(S)} = 0$ | $\hat{Q}_{0.999}^{(S)} \approx 2.2 \times 10^4$ <br> $\hat{\hat{Q}}_{0.999}^{(S)} \approx 3.2 \times 10^4$ <br> $\Delta \approx 0.47$ |
| "Naive model" | $\hat{\mu}_U \approx 3.040, \quad \hat{\sigma}_U \approx 1.509, \quad \hat{\theta} = 15.5$ <br> $\hat{\tau}_\mu^{(U)} \approx 0.193, \quad \hat{\tau}_\sigma^{(U)} \approx 0.137, \quad \hat{\tau}_\theta^{(U)} \approx 1.969$ <br> $\hat{\rho}_{\mu\sigma}^{(U)} = \hat{\rho}_{\mu\theta}^{(U)} = \hat{\rho}_{\sigma\theta}^{(U)} = 0$ | $\hat{Q}_{0.999}^{(U)} \approx 8.4 \times 10^3$ <br> $\hat{\hat{Q}}_{0.999}^{(U)} \approx 1.2 \times 10^4$ <br> $\Delta \approx 0.37$ |